\documentclass[iop]{emulateapj}

\usepackage{natbib,graphicx,amsmath}

\usepackage{color}
\usepackage{ulem}

\shorttitle{AXION AS A CDM CANDIDATE}
\shortauthors{NOH, HWANG \& PARK}

\newcommand{\bea}{\begin{eqnarray}}
\newcommand{\eea}{\end{eqnarray}}

\begin{document}

\title{Axion as a cold dark matter candidate: Proof to fully nonlinear order}
\author{Hyerim Noh${}^{1}$, Jai-chan Hwang${}^{2}$, Chan-Gyung Park${}^{3}$ }
\address{${}^{1}$Center for Large Telescope,
         Korea Astronomy and Space Science Institute, Daejon, Korea \\
         ${}^{2}$Department of Astronomy and Atmospheric Sciences, Kyungpook National University, Daegu, Korea \\
         ${}^{3}$Division of Science Education and Institute of Fusion Science, Chonbuk National University, Jeonju, Korea}

\begin{abstract}

We present a proof of the axion as a cold dark matter candidate to the fully nonlinear order perturbations based on Einstein's gravity. We consider the axion as a coherently oscillating massive classical scalar field without interaction. We present the fully nonlinear and exact, except for {\it ignoring} the transverse-tracefree tensor-type perturbation, hydrodynamic equations for an axion fluid in Einstein's gravity. We show that the axion has the characteristic pressure and anisotropic stress, the latter starts to appear from the second-order perturbation. But these terms do not directly affect the hydrodynamic equations in our axion treatment. Instead, what behaves as the effective pressure term in relativistic hydrodynamic equations is the perturbed lapse function and the relativistic result coincides exactly with the one known in the previous non-relativistic studies. The effective pressure term leads to a Jeans scale which is of the solar-system scale for conventional axion mass. As the fully nonlinear and relativistic hydrodynamic equations for an axion fluid coincide exactly with the ones of a zero-pressure fluid in the super-Jeans scale, we have proved the cold dark matter nature of such an axion in that scale.

\end{abstract}


\section{Introduction}

Although the status is still theoretical, the axion is one of the prime candidates for a cold dark matter (CDM) (Preskill et al. 1983; Abbott \& Sikivie 1983; Dine \& Fischler 1983). For theoretical aspects and the experimental search efforts of axion see Kim (1987), Kim \& Carosi (2010), Kawasaki \& Nakayama (2013), Graham et al. (2015).
Here we consider the axion as a coherently oscillating massive classical scalar field without interaction. Proof of such an axion as a CDM was shown to the linear perturbation order in non-relativistic context (Khlopov et al. 1985) and in Einstein's gravity (Nambu \& Sasaki 1990; Ratra 1991; Hwang 1997; Sikivie \& Yang 2009; Hwang \& Noh 2009). We have been studying the case to nonlinear perturbation orders and proved the CDM nature of the axion fluid in the super-Jeans scale up to the third order perturbations (Noh et al. 2013, 2015).

In this work now we prove the case to the fully nonlinear perturbation orders. We present the fully nonlinear and exact general relativistic hydrodynamic equations for an axion fluid in the axion-comoving gauge: see Equations (\ref{eq6})-(\ref{eq2}). We show that the axion has the characteristic pressure and anisotropic stress: see Equation (\ref{delta-N-p}). However, the axion nature appearing as an effective pressure term in the relativistic hydrodynamic equations comes from the perturbed lapse function, see Equation (\ref{eq4}), and the result coincides exactly with the one derived in the non-relativistic context, see Equation (\ref{Euler-eq}). This proves that the axion is indeed non-relativistic. We show that in the super-Jeans scale the axion fluid is the same as the zero-pressure fluid in Einstein's gravity to fully nonlinear order.

%
%
%
\section{Fully nonlinear and exact perturbation}
                                             \label{sec:FNLE}

We consider a spatially {\it flat} Robertson-Walker background with fully nonlinear perturbations; we include the cosmological constant in the background. As the metric we take
\bea
       d s^2
   &=& - a^2 \left( 1 + 2 \alpha \right) d \eta^2
       - 2 a \chi_i d \eta d x^i
   \nonumber \\
   & &
       + a^2 \left( 1 + 2 \varphi \right) \delta_{ij}
       d x^i d x^j,
   \label{metric-FNL}
\eea
where $\alpha$, $\varphi$ and $\chi_i$ are functions of space and time with arbitrary amplitudes; index of $\chi_i$ is raised and lowered by $\delta_{ij}$. The spatial part of the metric looks simple as we have {\it taken} a spatial gauge condition without losing any generality (Bardeen 1988; Hwang \& Noh 2013) and {\it ignored} the transverse-tracefree (TT) tensor-type spatial metric. The TT perturbation corresponds to the gravitational wave to the linear order. The inflation generated primordial TT perturbation is sufficiently weak not to contribute gravitational evolution of matter, and the TT part generated from the hydrodynamic evolution of matter to the second order perturbation is also weak (Hwang et al.\ 2017). In the post-Newtonian (PN) approximation the TT part is generated only from the 2.5PN order (Chandrasekhar \& Esposito 1970).

We introduce a perturbation variable $\kappa$ as the perturbed part of the trace of extrinsic curvature. In terms of the metric we have
\bea
   & & \kappa
       = 3 H \left( 1- {1 \over {\cal N}} \right)
   \nonumber \\
   & & \qquad
       - {1 \over {\cal N} (1 + 2 \varphi)}
       \left[ 3 \dot \varphi
       + {c \over a^2} \left( \chi^k_{\;\;,k}
       + {\chi^{k} \varphi_{,k} \over 1 + 2 \varphi} \right)
       \right],
   \label{eq1}
\eea
with ${\cal N}$ defined by the lapse function $N \equiv a {\cal N} = 1/\sqrt{-\widetilde g^{00}}$ (with $x^0 = \eta$), thus
\bea
   & & {\cal N} \equiv \sqrt{ 1 + 2 \alpha + {\chi^{k} \chi_{k} \over
       a^2 (1 + 2 \varphi)}}
      \equiv 1 + \delta {\cal N}.
\eea
These are exact. The fully nonlinear perturbation equations for fluids and fields in Einstein's gravity are presented in Hwang \& Noh (2013), Noh (2014), and Hwang et al. (2016). We set $c \equiv 1 \equiv \hbar$ unless stated otherwise.

%
%
%
\section{Axion}
                                             \label{sec:Axion}

As the axion we consider a massive minimally coupled scalar field with the energy-momentum tensor given as
\bea
       \widetilde T_{ab}
       = \widetilde \phi_{,a} \widetilde \phi_{,b}
       - {1 \over 2} \left( \widetilde \phi^{;c} \widetilde \phi_{,c}
       + m^2 \widetilde \phi^2 \right) \widetilde g_{ab},
   \label{Tab-MSF}
\eea
where tildes indicate the covariant quantities.

We decompose $\widetilde \phi ({\bf x}, t) \equiv \phi (t)  + \delta \phi ({\bf x}, t)$. To the background order we have a solution (Ratra 1991)
\bea
       \phi (t) = a^{-3/2} \left[ \phi_{+0} \sin{(mt)}
       + \phi_{-0} \cos{(mt)} \right],
   \label{BG-phi}
\eea
where $\phi_{+0}$ and $\phi_{-0}$ are the constant coefficients; $t$ is related to the conformal time $\eta$ as $c dt \equiv a d\eta$. As the axion coherently oscillates rapidly, we take the average over time scale of order $m^{-1}$, and strictly {\it ignore} ${H / m}$ higher order term (Ratra 1991). We have
\bea
       {H \over m} = 2.133 \times 10^{-28} h
       \left( { 10^{-5} {\rm eV} \over m } \right)
       \left( {H \over H_0} \right),
   \label{Hm}
\eea
where $H_0 \equiv 100h {\rm km} \, {\rm s}^{-1} {\rm Mpc}^{-1}$ represents the present Hubble parameter with $H \equiv \dot a/a$.

The oscillation time scale for axion is
\bea
       {\hbar \over m c^2} = 6.582 \times 10^{-11}
       \left( { 10^{-5} {\rm eV} \over m } \right)
       {\rm sec}.
   \label{oscillation-time}
\eea
The coherence time scale for oscillating axion which might be relevant for the laboratory detection is (Graham \& Rajendran 2013)
\bea
       {h \over m v^2} \sim 4 \times 10^{-4}
       \left( { 10^{-5} {\rm eV} \over m } \right)
       {\rm sec},
   \label{coherence-time}
\eea
with the galactic virial velocity $v \sim 10^{-3}c$; it is the time for the axion particle to cross the de Broglie wavelength, set by the galactic virial velocity of the axion particles.

%
%
%
\subsection{Perturbations}
                                             \label{sec:Perturbations}

We expand (Ratra 1991; Hwang 1997; Hwang \& Noh 2009; Noh et al. 2013, 2015)
\bea
       \delta \phi ({\bf x}, t)
       \equiv
       \delta \phi_+ ({\bf x}, t) \sin{(mt)}
       + \delta \phi_- ({\bf x}, t) \cos{(mt)},
   \label{ansatz}
\eea
where $\delta \phi_+$ and $\delta \phi_-$ have arbitrary amplitudes.

We take the average over the quadratic combinations of the scalar field in the right-hand-side of Equation \ (\ref{Tab-MSF}). This will give the smoothed $\widetilde T_{ab} = \langle \widetilde T_{ab} \rangle$ to fully nonlinear order in perturbation. From this we can read the fluid quantities given as
\bea
       \widetilde T_{ab} = \widetilde \mu \widetilde u_a \widetilde u_b
       + \widetilde p \left( \widetilde g_{ab}
       + \widetilde u_a \widetilde u_b \right)
       + \widetilde \pi_{ab},
   \label{Tab-fluid}
\eea
where $\widetilde \mu (\equiv \widetilde \varrho c^2)$, $\widetilde p$ and $\widetilde u_a$ are the energy density, pressure and fluid four-vector with $\widetilde u^c \widetilde u_c \equiv -1$. We will show that although a minimally coupled scalar field does not have the anisotropic stress our averaging process leads to a {\it non-vanishing} anisotropic stress, see Equation \ (\ref{delta-N-p}) and below it. We decompose the fluid quantities as
\bea
       \widetilde \mu \equiv \mu + \delta \mu, \quad
       \widetilde p \equiv p + \delta p, \quad
       \widetilde \pi_{ij} \equiv a^2 \Pi_{ij},
\eea
where indices of $\Pi_{ij}$ are raised and lowered by $\delta_{ij}$ as the metric. We define the perturbed order fluid three-velocity $v_i$ as
\bea
       \widetilde u_i
       \equiv a \gamma {v_i \over c}, \quad
       \gamma \equiv {1 \over \sqrt{ 1 -
       {v^k v_k \over c^2(1 + 2 \varphi)}}}.
   \label{u_i}
\eea
We raise and lower the index of $v_i$ by $\delta_{ij}$ as the metric; to the nonlinear order the $v_i$ should be determined order by order in perturbative manner.

To the background order we have (Ratra 1991)
\bea
      \mu = {1 \over 2} m^2 a^{-3} ( \phi_{+0}^2 + \phi_{-0}^2 ), \quad
      p = 0.
\eea
Thus, the background evolution of the axion behaves exactly the same as a pressureless ideal fluid (Preskill et al. 1983; Abbott \& Sikivie 1983; Dine \& Fischler 1983). This is generally true in the presence of both the spatial curvature and the cosmological constant $\Lambda$ in the background, and in the presence of other fluids and fields.

In the following we use the estimates of $\kappa$, $\chi_i$ and $\varphi$
\bea
      \hskip -.5cm
      \kappa \sim H \delta \propto H, \;\;\;
      \chi_i \sim {a^2 \over c\Delta} \kappa_{,i} \propto H, \;\;\;
      \varphi \sim {a^2 H^2 \over c^2 \Delta} \delta \propto H^2,
   \label{estimates}
\eea
which follow from Equations \ (\ref{eq6}), (\ref{eq3}) and (\ref{eq2}); we have ${\partial \over \partial t} \sim H$. As we ignore terms with $H/m$, this often implies ignoring $\varphi$ and $a^2 H^2/(k^2 c^2)$ terms ($k$ is the comoving wavenumber with $\Delta = - k^2$) when these terms involve the axion mass with $1/m^2$.

%
%
%
\subsection{Axion-comoving gauge}
                                \label{sec:Axion-comoving gauge}

We take the axion-comoving temporal gauge (hypersuface or slicing condition) as
\bea
       \widetilde T_i^{0,i} \equiv 0,
   \label{axion-CG}
\eea
to all perturbation orders. Together with our spatial gauge condition the axion-comoving gauge fixes the gauge mode completely, and each variable has a unique corresponding gauge-invariant counterpart to all perturbation orders (Bardeen 1988; Noh \& Hwang 2004; Hwang \& Noh 2013). Thus all variables in this gauge can be regarded as gauge-invariant but are still attached with the certain gauge condition.

The $\widetilde T^0_i$ to the fully nonlinear order is presented in Equation (4.11) of Noh et al. (2015). In a perturbative manner, we can show that the gauge condition in Equation \ (\ref{axion-CG}) leads to
\bea
       \widetilde T^0_i
       = - {m \over 2 a} a^{-3/2}
       \left( \phi_{+0} \delta \phi_{-,i}
       - \phi_{-0} \delta \phi_{+,i} \right),
\eea
thus
\bea
       {\delta \phi_+ \over \phi_{+0}}
       - {\delta \phi_- \over \phi_{-0}} = 0,
\eea
order by order to the fully nonlinear order. Thus the axion-comoving gauge implies $\widetilde T^0_i = 0$ to fully nonlinear order. In a perturbative manner we can show that $v_i = 0$ to all perturbation orders, thus $\widetilde u_a$ is the same as the normal-frame four-vector $\widetilde n_a$. In a more realistic situation with additional presence of other fluids (like baryon and radiation) the energy-momentum tensor in the above should be considered as the axion part of the energy-momentum tensor. The axion part can be combined with the multiple fluid formulation of the fully nonlinear perturbation theory presented in Hwang et al. (2016). In this case the four-vector $\widetilde u_a$ should be considered as the axion-fluid four vector.

From Equation \ (\ref{Tab-MSF}) we have
\bea
   & &
       \widetilde T^0_0
       = - \mu \left[ {{\cal N}^2 + 1 \over 2 {\cal N}^2}
       \left( 1 + 2 \Phi + \Phi^2 \right)
       - {\hbar^2 \Phi^{,i} \Phi_{,i}
       \over 2 m^2 c^2 a^2 (1 + 2 \varphi)} \right],
   \nonumber \\
   & &
       \widetilde T^i_j
       = - \mu \bigg[ {{\cal N}^2 - 1 \over 2 {\cal N}^2}
       \left( 1 + 2 \Phi + \Phi^2 \right) \delta^i_j
   \nonumber \\
   & & \qquad
       + {\hbar^2 \over m^2 c^2 a^2 ( 1 + 2 \varphi )}
       \left( {1 \over 2} \delta^i_j \Phi^{,k} \Phi_{,k}
       - \Phi^{,i} \Phi_{,j} \right) \bigg],
   \nonumber \\
   & &
       \widetilde T^0_i = 0 = \widetilde T^i_0,
   \label{Tab-axion}
\eea
where $\Phi \equiv {\delta \phi_+ /( a^{-3/2} \phi_{+0})}$; in Equations (\ref{oscillation-time})-(\ref{u_i}) and (\ref{estimates}), and from here we recover $c$ and $\hbar$.

%
%
%
\subsection{Equation of motion}
                            \label{sec:Equation-of-motion}

Using the equation of motion in Equation \ (4.7) of Noh et al.\ (2015), to the fully nonlinear order we can show
\bea
       \delta {\cal N}
       = {\hbar^2 \over 2 m^2 c^2 a^2 (1 + 2 \varphi)}
       {\Delta \Phi \over 1+\Phi }.
   \label{delta-N}
\eea
This relation follows from dividing Equation \ (4.7) of Noh et al.\ (2015) by $\phi_{\mp 0}$ and summing the two relations; we kept {\it only} the terms that are leading order in ${\Delta \over m^2 a^2} \Phi$.

%
%
%
\subsection{Fluid quantities}
                                 \label{sec:Fluid-quantities}

From Equations \ (\ref{Tab-fluid}) and (\ref{Tab-axion}) we can read the fluid quantities
\bea
   & &
       \delta \equiv {\delta \varrho \over \varrho}
       = 2 \Phi + \Phi^2
       - {\hbar^2 \Delta( 2 \Phi + \Phi^2) \over 4 m^2 c^2 a^2 (1 + 2 \varphi)},
   \nonumber \\
   & &
       \delta p
       =
       - \varrho {\hbar^2 \Delta( 2 \Phi + \Phi^2) \over 4 m^2 a^2 (1 + 2 \varphi)},
   \nonumber \\
   & &
       \Pi_{ij} = \varrho {\hbar^2 \over m^2 a^2}
       \left( \Phi_{,i} \Phi_{,j}
       - {1 \over 3} \delta_{ij} \Phi^{,k} \Phi_{,k} \right),
   \label{p-Phi}
\eea
valid to fully nonlinear order. For our purpose of expressing $\delta {\cal N}$ in terms of $\delta$, we need the expression in terms of $\delta$ only to the zeroth order in ${\Delta \over m^2 a^2} \Phi$, thus
\bea
       \Phi = -1 \pm \sqrt{1 + \delta}.
   \label{Phi-delta}
\eea
Then, from Equations \ (\ref{delta-N}) and (\ref{p-Phi}) we have
\bea
   & &
       \delta {\cal N}
       = {\hbar^2 \over 2 m^2 c^2 a^2 (1 + 2 \varphi)}
       {\Delta \sqrt{\widetilde \varrho}
       \over \sqrt{\widetilde \varrho}},
   \nonumber \\
   & &
       \delta p
       = - {\hbar^2 \Delta \widetilde \varrho \over 4 m^2 a^2 (1 + 2 \varphi)},
   \nonumber \\
   & &
       \Pi_{ij} = {\hbar^2 \over 4 m^2 a^2 \widetilde \varrho}
       \left( \widetilde \varrho_{,i}
       \widetilde \varrho_{,j}
       - {1 \over 3} \delta_{ij} \widetilde \varrho^{,k}
       \widetilde \varrho_{,k} \right).
   \label{delta-N-p}
\eea
These are the perturbed lapse function, pressure and anisotropic stress generated by the axion fluid. Later we will show that in our approximation of ignoring the $H/m$-term and keeping the terms only to the leading order in ${\Delta \over m^2 a^2} \Phi$, the pressure and anisotropic stress terms disappear and only the perturbed lapse function affects the hydrodynamic equations, see Equation \ (\ref{eq4}). These three terms are related to each other, see Equation (\ref{delta-N-p-pi-2}).

It is curious to see the appearance of the anisotropic stress in the case of axion after time averaging. The anisotropic stress of the minimally coupled scalar field vanishes in the covariant level due to the energy-frame condition we have taken (without losing any generality) in Equation (\ref{Tab-fluid}); i.e., we have
\bea
   & & \widetilde \mu
       = \widetilde T_{ab} \widetilde u^a \widetilde u^b, \quad
       \widetilde p = {1 \over 3} \widetilde T_{ab} \widetilde h^{ab}, \quad
       \widetilde \pi_{ab}
       = \widetilde T_{cd} \widetilde h^c_a \widetilde h^d_b
       - \widetilde p \widetilde h_{ab},
   \nonumber \\
   & &
       \widetilde q_a = - \widetilde T_{cd} \widetilde u^c \widetilde h^d_a
       \equiv 0,
\eea
with $\widetilde h_{ab} \equiv \widetilde g_{ab} + \widetilde u_a \widetilde u_b$ the spatial projection tensor. For the scalar field, we have
\bea
   & & \widetilde q_a = - \widetilde h^b_a \widetilde \phi_{,b}
       \widetilde \phi_{,c} \widetilde u^c.
\eea
The energy-frame condition $(\widetilde q_a \equiv 0$) gives $\widetilde h^b_a \widetilde \phi_{,b} = 0$ with $\widetilde \pi_{ab} = 0$ and
\bea
   \hskip -.5cm
   \widetilde \mu
       = {1 \over 2} ( \widetilde \phi_{,c} \widetilde u^c )^2
       + {1 \over 2} m^2 \widetilde \phi^2, \quad
       \widetilde p
       = {1 \over 2} ( \widetilde \phi_{,c} \widetilde u^c )^2
       - {1 \over 2} m^2 \widetilde \phi^2.
\eea
Thus a minimally coupled scalar field has no anisotropic stress in the covariant level.

From $\widetilde h^b_i \widetilde \phi_{,b} = \widetilde \phi_{,i} + \widetilde u_i \widetilde \phi_{,b} \widetilde u^b = 0$ we have
\bea
   & & \widetilde T^0_i = - \left( \widetilde \mu + \widetilde p \right)
       \widetilde u^0 {\widetilde \phi_{,i} \over
       \widetilde \phi_{,b} \widetilde u^b}.
\eea
If we further impose the comoving gauge condition in Equation (\ref{axion-CG}) we have $\widetilde \phi_{,i} = 0$ order by order, thus $\widetilde \phi = \phi(t)$ and $\widetilde u_i = 0 = \widetilde T^0_i$ to all perturbation orders. Thus, in the comoving gauge we have
\bea
   & & \widetilde T^0_0 = - \widetilde \mu, \quad
       \widetilde T^0_i = 0, \quad
       \widetilde T^i_j = \widetilde p \delta^i_j.
\eea
As we notice in Equation (\ref{Tab-axion}), in the axion case the anisotropic stress somehow survives through the temporal averaging even in the axion-comoving gauge.

%
%
%
\subsection{Complete equations}
                                  \label{sec:CE}

In order to derive the hydrodynamic equations we can use the fully nonlinear perturbation equations with the fluid quantities in Equation \ (\ref{delta-N-p}) and $v_i = 0$. The Arnowitt-Deser-Misner (ADM) energy conservation and the trace of ADM propagation equations in Equations \ (16) and (9) of Hwang et al.\ (2016), respectively, give
\begin{widetext}
\bea
   & &
       \dot \delta - \kappa
       + {c \over a^2 (1+2\varphi)}
       \delta_{,i} \chi^i
       - \delta \kappa
       = 0,
   \label{eq6} \\
   & &
       \dot \kappa
       + 2 H \kappa
       - 4 \pi G \delta \varrho
       + {c \over a^2 (1 + 2 \varphi)} \kappa_{,i} \chi^i
       - {1 \over 3} \kappa^2
              - {c^2 \over a^4 (1 + 2\varphi)^2}
       \bigg\{
       {1 \over 2} \chi^{i,j} \left( \chi_{i,j} + \chi_{j,i} \right)
              - {1 \over 3} \chi^i_{\;\;,i} \chi^j_{\;\;,j}
              \nonumber \\
     & & \qquad
          - {4 \over 1 + 2 \varphi} \left[
       {1 \over 2} \chi^i \varphi^{,j} \left(
       \chi_{i,j} + \chi_{j,i} \right)
       - {1 \over 3} \chi^i_{\;\;,i} \chi^j \varphi_{,j} \right]
              + {2 \over (1 + 2 \varphi)^2} \left(
       \chi^{i} \chi_{i} \varphi^{,j} \varphi_{,j}
       + {1 \over 3} \chi^i \chi^j \varphi_{,i} \varphi_{,j} \right) \bigg\}
   \nonumber \\
   & & \qquad
       = - {c^2 \Delta {\cal N} \over a^2 (1 + 2 \varphi)}
       = - {\hbar^2 \Delta \over 2 m^2 a^4 (1 + 2 \varphi)}
       {\Delta \sqrt{1 + \delta} \over \sqrt{1 + \delta}}.
   \label{eq4}
\eea
In order to close the system we need the following two constraint equations. The ADM momentum constraint and energy constraint equations in Equations \ (8) and (7) of Hwang et al.\ (2016), respectively, give
\bea
   & &
       {2 \over 3} \kappa_{,i}
       + {c \over 2 a^2 ( 1 + 2 \varphi )}
       \left( \Delta \chi_i
       + {1 \over 3} \chi^k_{\;\;,ik} \right)
       =
       {c \over a^2 ( 1 + 2 \varphi)} \bigg\{
       - {\varphi_{,j} \over 1 + 2 \varphi}
       \Big[ {1 \over 2} \left( \chi^{j}_{\;\;,i} + \chi_i^{\;,j} \right)
       \nonumber \\
    & & \qquad
       - {1 \over 3} \delta^j_i \chi^k_{\;\;,k} \Big]
          - {\varphi^{,j} \over (1 + 2 \varphi)^2}
       \left( \chi_{i} \varphi_{,j}
       + {1 \over 3} \chi_{j} \varphi_{,i} \right)
             + {1 \over 1 + 2 \varphi} \nabla_j
       \Big(
       \chi^{j} \varphi_{,i}
       + \chi_{i} \varphi^{,j}
       - {2 \over 3} \delta^j_i \chi^{k} \varphi_{,k} \Big)
       \bigg\},
   \label{eq3} \\
   & &
       {c^2 \Delta \varphi \over a^2 (1 + 2 \varphi)^2}
       + 4 \pi G \delta \varrho
       + H \kappa
       = {1 \over 6} \kappa^2
       + {3 \over 2} {c^2 \varphi^{,i} \varphi_{,i} \over a^2 (1 + 2 \varphi)^3}
             - {c^2 \over 4 a^4 (1 + 2 \varphi)^2}
       \bigg\{
       {1 \over 2} \chi^{i,j} \left( \chi_{i,j} + \chi_{j,i} \right)
       - {1 \over 3} \chi^i_{\;\;,i} \chi^j_{\;\;,j}
   \nonumber \\
   & & \qquad
       - {4 \over 1 + 2 \varphi} \left[
       {1 \over 2} \chi^i \varphi^{,j} \left(
       \chi_{i,j} + \chi_{j,i} \right)
       - {1 \over 3} \chi^i_{\;\;,i} \chi^j \varphi_{,j} \right]
             + {2 \over (1 + 2 \varphi)^2} \left(
       \chi^{i} \chi_{i} \varphi^{,j} \varphi_{,j}
       + {1 \over 3} \chi^i \chi^j \varphi_{,i} \varphi_{,j} \right) \bigg\}.
   \label{eq2}
\eea
\end{widetext}
Equations (\ref{eq6})-(\ref{eq2}) are valid to the fully nonlinear order and are exact. For a zero-pressure fluid these were presented in Hwang \& Noh (2013) ignoring the vector-type perturbation. Here we include the vector perturbation.

We find that the axion nature only appears in the right-hand-side of Equation \ (\ref{eq4}). We note that the pressure and anisotropic stress terms from the axion nature in Equation \ (\ref{delta-N-p}) disappear in the above equations due to our estimates in Equation \ (\ref{estimates}). In the above hydrodynamic equations the axion nature appears only through the perturbed lapse function $\delta {\cal N}$ which directly follows from the equation of motion, see Equation \ (\ref{delta-N}); the $\varphi$ term in the right-hand-side of Equation \ (\ref{eq4}), and similarly the ones in Equations \ (\ref{Tab-axion})-(\ref{delta-N-p}), can be ignored due to the estimate in Equation \ (\ref{estimates}) and below it. The momentum conservation equation in Equation \ (15) or (17) of Hwang et al.\ (2016) gives
\bea
      \widetilde p_{,i}
      + \left( \widetilde \mu + \widetilde p \right) {\cal N}_{,i}
      = - {\hbar^2 \over 4 m^2 a^2 (1 + 2 \varphi)}
      \left( {\widetilde \varrho^{,j} \widetilde \varrho_{,i}
      \over \widetilde \varrho} \right)_{,j},
   \label{delta-N-p-pi}
\eea
which is consistent with Equation \ (\ref{delta-N-p}). To the linear order we have $\delta p/\mu = - \delta {\cal N}$.

%
%
%
\subsection{Newtonian correspondence}
                                   \label{sec:correspondence}

In order to compare the relativistic equations with the Newtonian fluid equations, we {\it identify} $\delta$ as the density contrast ${\delta \varrho / \varrho}$, and $\kappa$ as the divergence of velocity perturbation
\bea
       \kappa \equiv - {1 \over a} \nabla \cdot {\bf u}
       \equiv - {\Delta \over a} u.
   \label{kappa-u}
\eea
Equations (\ref{eq6}) and (\ref{eq4}) can be identified as the energy and the momentum conservation equations, respectively (Noh \& Hwang 2004). The variable $\chi_i$ can be related to $\kappa$ using Equation \ (\ref{eq3}), and $\varphi$ can be determined using Equation \ (\ref{eq2}), both in perturbative manner. Equations (\ref{eq6})-(\ref{eq2}) provide closed hydrodynamic equations for an axion fluid valid to fully nonlinear order (except that we ignored the transverse-tracefree tensor-type perturbation in the spatial metric). To the third order, for example, we need Equations \ (\ref{eq6}) and (\ref{eq4}) to the third order, Equation \ (\ref{eq3}) to the second order, and Equation \ (\ref{eq2}) to the linear order.

%
%
%
\subsection{Weak gravity limit}
                               \label{sec:WG}

For negligible $\varphi$ term compared with unity (weak gravity limit), Equations \ (\ref{eq6}), (\ref{eq4}) and (\ref{eq3}) become simplified a lot to {\it coincide} exactly with Newtonian equations. Decomposing $\chi_i \equiv \chi_{,i} + \chi^{(v)}_i$ with $\chi^{(v)i}_{\;\;\;\;\;\;,i} \equiv 0$, Equation \ (\ref{eq3}) gives
\bea
       \kappa + {\Delta\over a^2} \chi = 0, \quad
       \chi^{(v)}_i = 0.
    \label{varphi-lin-eq}
\eea
Thus $\chi_i = \chi_{,i} = a u_i/c$; we have $u_i \equiv u_{,i} + u_i^{(v)}$ with $u^{(v)i}_{\;\;\;\;\;\;,i} \equiv 0$, and $\chi^{(v)}_i = 0$ implies $u^{(v)}_i = 0$. Equations (\ref{eq6}) and (\ref{eq4}) give the energy and the momentum conservation equations
\bea
   & &
       \dot \delta
       + {1 \over a} \nabla \cdot {\bf u}
       + {1 \over a} \nabla \cdot \left( \delta {\bf u} \right)
       = 0.
          \label{continuity-eq} \\
   & &
       {1 \over a} \nabla \cdot \Big( \dot {\bf u}
       + H {\bf u} \Big)
       + 4 \pi G \varrho \delta
       + {1 \over a^2} \nabla \cdot \left( {\bf u} \cdot \nabla {\bf u} \right)
   \nonumber \\
   & & \qquad
       = {\hbar^2 \Delta \over 2 m^2 a^4} {\Delta \sqrt{1 + \delta}
       \over \sqrt{1 + \delta}}.
   \label{Euler-eq}
\eea
So, we have the Newtonian correspondence (Noh \& Hwang 2004; Hwang \& Noh 2013; Hwang et al.\ 2014) except the axion pressure term. The axion pressure term is exactly the same as the one derived in non-relativistic limit with the Schr\"odinger equation (Madelung 1927; Chavanis 2012; Uhlemann et al.\ 2014; Marsh 2015). Combining Equations \ (\ref{continuity-eq}) and (\ref{Euler-eq}) we obtain
\bea
   & &
       \ddot \delta + 2 H \dot \delta
       - 4 \pi G \varrho \delta
       + {1 \over a^2} \left[ a \nabla \cdot \left( \delta {\bf u} \right)
       \right]^{\displaystyle\cdot}
       - {1 \over a^2} \nabla \cdot \left( {\bf u}
       \cdot \nabla {\bf u} \right)
   \nonumber \\
   & & \qquad
       = - {\hbar^2 \Delta \over 2 m^2 a^4} {\Delta \sqrt{1 + \delta}
       \over \sqrt{1 + \delta}}.
   \label{ddot-delta-eq}
\eea
The left-hand-side coincides exactly with the Newtonian equation valid to fully nonlinear order (Peebles 1980; Noh \& Hwang 2004; Hwang \& Noh 2013). For a zero-pressure fluid the Newtonian equations are closed to the second order. The pure Einstein's gravity corrections appear from the third-order and {\it all} terms involve $\varphi$, see Equations (\ref{eq6})-(\ref{eq2}). In the weak field limit Equation \ (\ref{eq2}) gives
\bea
   & &
       c^2 {\Delta \over a^2} \varphi
       = - 4 \pi G \delta \varrho
       + H {1 \over a} \nabla \cdot {\bf u}
   \nonumber \\
   & & \qquad
       + {1 \over 4 a^2} \left[ \left( \nabla \cdot {\bf u} \right)^2
       - u^{,ij} u_{,ij} \right],
\eea
where $u_{,ij} = \nabla_i \nabla_j u$. This can be regarded as a relation determining $\varphi$ the curvature perturbation in the (axion-)comoving gauge (Hwang et al.\ 2014).

%
%
%
\subsection{Axionic character}
                                   \label{sec:Jeans}

Comparing the gravity term with the pressure term to the linear order we have the Jeans scale given as (Nambu \& Sasaki 1990; Sikivie \& Yang 2009; Hwang \& Noh 2009)
\bea
      \hskip -.5cm
      \lambda_{J_a} \equiv {2 \pi a \over k_{J_a}}
      = \sqrt{ {\pi \hbar \over m} \sqrt{\pi \over G \varrho} }
      \sim 5.4 \times 10^{14} \textrm{cm} \sqrt{{10^{-5} \textrm{eV} \over m h}}.
\eea
Beyond this axion-Jeans scale Equations \ (\ref{eq6})-(\ref{eq2}) coincide exactly with the relativistic hydrodynamic equations of a zero-pressure fluid to fully nonlinear order. This stability scale was first known in non-relativistic context (Khlopov et al.\ 1985; Bianchi et al.\ 1990). Hu et al.\ (2000) interpret it as a de Broglie wavelength with
\bea
      \lambda
      \sim {h \over m v_g}
      \sim {h \over m \lambda/t_g}
      \sim {h \over m \lambda \sqrt{G \varrho}},
\eea
where $v_g \sim \lambda/t_g$ and $t_g \sim 1/\sqrt{G \varrho}$ the gravitational time scale. In this sense, stability below $\lambda_{J_a}$ is due to the uncertainty principle.

The only correction term in the hydrodynamic equations arising from the axion nature is
\bea
       c^2 {\Delta \over a^2} {\cal N}
       = {\hbar^2 \Delta \over 2 m^2 a^4} {\Delta \sqrt{1 + \delta}
       \over \sqrt{1 + \delta}}
       = {\hbar^2 \Delta \over 2 m^2 a^4} {\Delta \sqrt{\widetilde \varrho}
       \over \sqrt{\widetilde \varrho}},
   \label{axion-pressure}
\eea
which is often known as the quantum pressure term. Thus, even to the fully nonlinear and exact order in Einstein's gravity, this effective axion pressure term coincides exactly with the non-relativistic one obtained from Schr\"odinger equation in the  Minkowski space-time (Madelung 1927; Dalfovo et al.\ 1999; Pethick \& Smith 2002;  Pitaevskii \& Stringari 2003; Barcelo et al.\ 2005). It may not be a surprise that in the non-relativistic limit we recover the same results as expected from the Schr\"odinger equation; we used the Klein-Gordon equation which is a relativistic version of Schr\"odinger equation for bosons.

In our relativistic treatment, the anisotropic stress arising from the axion nature in Equation (\ref{delta-N-p}) does not directly affect the hydrodynamic equations; in this context the role of pressure term is the same. It is the lapse function caused by the pressure and anisotropic stress which represent the effective pressure term in the momentum conservation equation. From Equations (\ref{delta-N-p}) and (\ref{delta-N-p-pi}) we can show the relation among these three variables
\bea
   & & \widetilde p_{,i}
      + \left( \widetilde \mu + \widetilde p \right) {\cal N}_{,i}
      = - {1 \over 1 + 2 \varphi} \Pi^j_{i,j}
   \nonumber \\
   & & \qquad
      - {2 \over 3} {1 \over 1 + 2 \varphi}
      \left[ \left( 1 + 2 \varphi \right)
      \left( \delta p + \widetilde \mu \delta {\cal N} \right)
      \right]_{,i}.
   \label{delta-N-p-pi-2}
\eea
The anisotropic stress appears from the second order in perturbation.

Roles of the axionic pressure terms in the linear and nonlinear evolution stage of the large-scale cosmic structure, in the case of extreme low-mass axion are studied by Hui et al (2017) and Mocz et al (2017).

%
%
%
\section{Discussion}
                              \label{sec:Discussion}

In this work we have shown that the axion as a massive coherently oscillating scalar field behaves as non-relativistic zero-pressure fluid in the super-Jeans scale. The effectivr pressure term of the axion fluid is the same as the one known in the non-relativistic analysis (Madelung 1927; Dalfovo et al.\ 1999; Pethick \& Smith 2002;  Pitaevskii \& Stringari 2003; Barcelo et al.\ 2005; Chavanis 2012). Here we have treated the axion as a massive scalar field without any self-interaction. Interaction terms, if becomes important, may cause qualitatively different changes which are beyond the scope of this work. Although we presented our proof in a single axion field case, the extension to realistic multi-component situations with additional presence of baryon and radiation, etc.\ is trivial; in the hydrodynamic case, see Hwang et al.\ (2016).

In a zero-pressure fluid, our Equations (\ref{eq6})-(\ref{eq2}) show that the pure Einstein's gravity correction terms start to appear from the third order, and all the correction terms involve $\varphi$ (Hwang \& Noh 2013). The leading nonlinear power spectra of the density and velocity perturbations show that the pure Einstein's gravity correction terms appearing in the third order are entirely negligible compared with the relativistic/Newtonian power spectra in all scales in the context of current concordance cosmology (Jeong et al.\ 2011). Thus, in the zero-pressure medium (with the cosmological constant) the Newtonian analysis is quite reliable at least up to weakly nonlinear stages. In the super-Jeans scale, therefore, we have proved that the axion behaves as a CDM (zero-pressure fluid) independently of whether the gravity is relativistic or Newtonian. In the relativistic case, Equations \ (\ref{eq6})-(\ref{eq2}) are the fully nonlinear equations for an axion fluid in the axion-comoving gauge.

%
%
\acknowledgments

We wish to thank the anonymous referee for many constructive comments and suggestions. H.N.\ was supported by National Research Foundation of Korea funded by the Korean Government (No.\ 2015R1A2A2A01002791).
J.H.\ was supported by Basic Science Research Program through the National Research Foundation (NRF) of Korea funded by the Ministry of Science, ICT and future Planning (No.\ 2016R1A2B4007964).
C.G.P.\ was supported by Basic Science Research Program through the National Research Foundation of Korea (NRF) funded by the Ministry of Education (No. 2017R1D1A1B03028384).

%
%


\end{document}